\documentclass[final]{ias2}

\usepackage{graphicx}
\usepackage{multirow}
\usepackage{array}
\usepackage{amsmath}
\usepackage{hyperref}
\usepackage{float}
\def\be{\begin{equation}}
\def\ee{\end{equation}}
\def\ba{\begin{array}}
\def\ea{\end{array}}
\def\bea{\begin{eqnarray}}
\def\eea{\end{eqnarray}}
\begin{document}


\title{Investigation of $\Omega_{ccb}$ and $\Omega_{cbb}$ baryons in Regge phenomenology}

\author[]{J\lowercase{uhi} O\lowercase{udichhya$^{a}$}}
\email{juhioudichhya01234@gmail.com}
\author[]{K\lowercase{eval} G\lowercase{andhi$^{b}$}}
\author[]{\lowercase{and} A\lowercase{jay} K\lowercase{umar}  R\lowercase{ai$^{a}$}}
\address[sin]{Department of Physics, Sardar Vallabhbhai National Institute of Technology, Surat, Gujarat-395007, India}

\address[]{Department of Computer Sciences and Engineering,
	Institute of Advanced Research, Gandhinagar, Gujarat 382426, India}

\begin{abstract}

	Triply heavy baryons with quark content $ccb$ and $cbb$ are investigated within the framework of Regge phenomenology. With the assumption of linear Regge trajectories, we have extracted the relations between Regge parameters and baryon masses. Using these relations, we compute the ground state masses of $\Omega_{ccb}$ and $\Omega_{cbb}$ baryons. Further, the Regge slopes and intercepts are estimated for these baryons to obtain the excited state masses in the $(J,M^{2})$ and $(n,M^{2})$ planes. \textbf{Also, using the obtained results we calculate the other properties like magnetic moment and radiative decay width of these triply heavy baryons.} We compare our evaluated results with those obtained by the other theoretical approaches, and our results show a general agreement with them. The present study and our predictions will provide significant clues for future experimental research of these unseen triply heavy baryons.

%
%

\end{abstract}



\maketitle


\section{Introduction}

 The internal dynamics of heavy baryons and the nature of their excitation mechanisms can be revealed by studying their excited states. After, the significant experimental progress in discovering the singly and doubly heavy baryons \cite{EPJCCas_c,LHCb2020Cas_c,Bellecas_c(2970),LHCbOmega_c,LHCbOmega,Cas_cc,LHCbcas_bc}, now the next significant step ahead is the identification of triply heavy baryons. 
 Triply heavy baryons are composed of three heavy quarks ($c$ or $b$).  There are four baryons having three heavy quarks, namely $\Omega_{ccc}$, $\Omega_{bbb}$, $\Omega_{ccb}$, and $\Omega_{cbb}$. The mass spectra of $\Omega_{ccc}$ and $\Omega_{bbb}$ have been already calculated in our previous work \cite{JuhiOmega,Juhibottom}. 
 Since none of these triply heavy baryons have yet been reported experimentally, their production is highly challenging.
 Baranov \textit{et. al.} predicted that the baryon with three charm quarks might not be seen in $e^{+}e^{-}$ collisions and the predictions for the  baryon composed of three bottom quarks would be substantially worse \cite{Baranov2004}.  First estimations of the production cross-section of the baryons with three heavy quarks at the LHC were assessed in the Refs. \cite{Gomshi2003,Gomshi2005,Gomshi2006}. In 2011, it was estimated that around $10^{4}-10^{5}$ events of $\Omega_{ccc}$ and $\Omega_{ccb}$ baryons, could be accumulated for 10 $fb^{-1}$ integrated luminosity at LHC \cite{Y.Q. Chen2011}. In Refs. \cite{Y. Liu2015,J. Zhao2017,S. Cho2017} certain \textbf{analyses} of the production rate of multi-charmed hadrons in heavy-ion collisions at high energies are discussed. This indicates that there is a good chance of triply charmed baryon to be observed at LHC.

 \textbf{In the last} few years, many authors have studied the properties of triply heavy baryons using different theoretical approaches. In a recent study \cite{Faustov2022}, the authors obtained the ground as well as excited state masses of triply heavy baryons by employing the relativistic quark model based on the quark-diquark picture in the quasipotential approach in Quantum chromodynamics. Shah et. al. \cite{Shah2018,Zalak2018} determined the radial and orbital excited state masses of the triply heavy baryons within the framework of the hypercentral constituent quark model with the addition of the first order correction in the potential term of the Hamiltonian. Further, the Regge trajectories for these baryons are also drawn. The authors of Ref. \cite{G.Yang2020} computed the mass spectra of triply heavy baryons using the constituent quark model.
 The Gaussian expansion method is used to solve the non-relativistic three-body bound state problem. The nonrelativistic quark model with the harmonic oscillator wave functions was used to calculate the triply heavy baryon masses in the Ref. \cite{Roberts2008}. In Ref. \cite{Z.G.Wang2012} the authors computed the masses of triply heavy baryons and \textbf{made} reasonable predictions using QCD sum rules. The ground state $\frac{1}{2}^{+}$ and $\frac{3}{2}^{+}$ masses of $\Omega_{ccb}$ and $\Omega_{cbb}$ baryons are evaluated by employing the QCD sum rules in the Refs. \cite{T. M. Aliev2013,T. M. Aliev2014}. 
 
 Also, there have been other theoretical mass determinations including non-relativistic quark model \cite{B. Patel20009,Vijande2015}, relativistic quark model \cite{S. Migura2006,A. P. Martynenko2008}, QCD sum rule \cite{J. R. Zhang2009},  continuum approach to QCD based on the Dyson-Schwinger equations \cite{P.-L. Yin2019,S.-x. Qin2019}, lattice gauge theories \cite{M. Padmanath2014,Padmanath2018}, Bag model \cite{A. Bernotas2009,P. Hasenfratz1980}, and potential non-relativistic QCD (pNRQCD) \cite{N. Brambilla2005,N. Brambilla2010}.  Since various theoretical approaches \textbf{give} different predictions and a wide range of mass \textbf{differences} can be seen in the outcomes of different models for these triply heavy baryons. Therefore, more comparisons
 of computations is required in the family of triply heavy $\Omega$, which motivates us to study these baryons. 
 In the present work, the spectroscopic study of triply heavy baryons with $ccb$ and $cbb$ quark content is done. The ground as well as excited state masses are obtained within the framework of Regge phenomenology. With the assumption of quasi-linear Regge trajectories, the relations between Regge parameters and baryon masses have been derived. With the help of  these extracted relations the ground state masses of these unseen triply heavy baryons \textbf{are estimated}. Further, the Regge parameters are computed to obtain the orbital and radial excited state masses of $\Omega_{ccb}$ and $\Omega_{cbb}$ baryons.
 
 The present article is arranged as follows. After briefing the various theoretical approaches studies the triply heavy baryons, in Sec 2 we explain the complete Regge theory and \textbf{derive} certain relations in terms of baryon masses, Regge slopes, and intercepts. The ground state masses of these unseen triply heavy baryons; $\Omega_{ccb}$ and $\Omega_{cbb}$ have been evaluated with the aid of the extracted relations. After that, Regge parameters are estimated and with the help of the values of Regge slopes and intercepts, the excited state masses are obtained in the $(J,M^{2})$ and $(n,M^{2})$ planes. \textbf{Sec. 3 deals with the magnetic moment followed by the calculation of the radiative decay widths of these baryons in Sec. 4.} Sec. 5 provides a detailed description of our obtained results. Finally, we concluded our work in Sec. 6. 

\section{Theoretical Framework}

In 1978, Nambu put forth the simplest hypothesis to account for the linear Regge trajectories  \cite{Nambu1974,Nambu1979}.
The Regge theory has been successfully used in our previous work to obtain the mass spectra of light as well as heavy baryons \cite{JuhiOmega,Juhibottom,physica,Juhilight}. We use the same phenomenology in the present work to study the triply heavy baryons. \textbf{Since, there is no experimental evidence is found in the field of triply heavy baryons, but the various theoretical approaches and the predicted results shows that the baryons ranging from light to singly, doubly, and triply heavy follows the linear behaviour \cite{Faustov2022,ChandniOmega,AmeeIJMPA2022,AmeeIJMPA,ZalakUniverse2021}.Hence, we have taken the quasi-linear Regge trajectories for these triply heavy baryons.}
The general form of linear Regge trajectories can be expressed as \cite{JuhiOmega,Juhibottom,Wei2008}, 

\begin{equation}
	\label{eq:1}
	J = \alpha(M) = a(0)+\alpha^{'} M^{2} ,
\end{equation}
\textbf{where $a(0)$ and $\alpha^{'}$ denotes the intercept and slope
of the Regge trajectory respectively.} These Regge intercepts and slopes for
different flavors of a baryon multiplet can be related by two relations:
\\
the additivity of intercepts;
\begin{equation}
	\label{eq:2}
	a_{ppr}(0) + a_{qqr}(0) = 2a_{pqr}(0) ,	
\end{equation}
the additivity of inverse slopes;
\begin{eqnarray}
	\label{eq:3}
	\frac{1}{{\alpha^{'}}_{ppr}} + \frac{1}{{\alpha^{'}}_{qqr}} = \frac{2}{{\alpha^{'}}_{pqr}} ,	
\end{eqnarray}
where $p, q,$ and  $r$ represent the quark flavors.
These two relations were derived in a $q\overline{q}$ string picture of hadrons \cite{Add4} and are generalised for baryons as well by assuming the baryon as a quark-diquark string object \cite{Add5,Add3,Add6,Add7}. Also, the ref. \cite{Add3} showed that the additivity of inverse Regge slopes is consistent with the formal chiral and heavy quark limits for both mesons and baryons. \textbf{For a} detailed description of these two relations, see \cite{Add5,Wei2008} and references therein.


Now using Eqs. (\ref{eq:1}) and (\ref{eq:2}) together with Eq. (\ref{eq:3}), we get two pairs of solutions expressed in terms of slope ratios and baryon masses,

\begin{eqnarray}
	\label{eq:4} \nonumber
	\frac{\alpha^{'}_{qqr}}{\alpha^{'}_{ppr}}&=&\frac{1}{2M^{2}_{qqr}}\times[(4M^{2}_{pqr}-M^{2}_{ppr}-M^{2}_{qqr}) \\
	&\pm&\sqrt{{{(4M^{2}_{pqr}-M^{2}_{ppr}-M^{2}_{qqr}})^2}-4M^{2}_{ppr}M^{2}_{qqr}}],
\end{eqnarray}
and, 
\begin{eqnarray} \nonumber
	\label{eq:5}
	\frac{\alpha^{'}_{pqr}}{\alpha^{'}_{ppr}}&=&\frac{1}{4M^{2}_{pqr}}\times[(4M^{2}_{pqr}+M^{2}_{ppr}-M^{2}_{qqr})\\
	&\pm&\sqrt{{{(4M^{2}_{pqr}-M^{2}_{ppr}-M^{2}_{qqr}})^2}-4M^{2}_{ppr}M^{2}_{qqr}}].
\end{eqnarray}
The above obtained Eq. (\ref{eq:4}) can also be expressed as,
\begin{equation}
	\label{eq:6}
	\frac{\alpha^{'}_{qqr}}{\alpha^{'}_{ppr}}=	\frac{\alpha^{'}_{kkr}}{\alpha^{'}_{ppr}}\times	\frac{\alpha^{'}_{qqr}}{\alpha^{'}_{kkr}} ,
\end{equation}	
here $k$ can be any quark flavor. Thus we have \cite{Wei2008},
\begin{eqnarray}
	\label{eq:7}
	\nonumber
	\frac{[(4M^{2}_{pqr}-M^{2}_{ppr}-M^{2}_{qqr})+\sqrt{{{(4M^{2}_{pqr}-M^{2}_{ppr}-M^{2}_{qqr}})^2}-4M^{2}_{ppr}M^{2}_{qqr}}]}{2M^{2}_{qqr}} \\  \hspace{-2cm}
	=\frac{[(4M^{2}_{pkr}-M^{2}_{ppr}-M^{2}_{kkr})+\sqrt{{{(4M^{2}_{pkr}-M^{2}_{ppr}-M^{2}_{kkr}})^2}-4M^{2}_{ppr}M^{2}_{kkr}}]/2M^{2}_{kkr}}{[(4M^{2}_{qkr}-M^{2}_{qqr}-M^{2}_{kkr})+\sqrt{{{(4M^{2}_{qkr}-M^{2}_{qqr}-M^{2}_{kkr}})^2}-4M^{2}_{qqr}M^{2}_{kkr}}]/2M^{2}_{kkr}} . 
\end{eqnarray}
\\
This is the significant relationship we have derived between the masses of different flavors of baryons. This relation can be used to estimate the mass of any baryon state if all other masses are known. In the present work the ground state masses of $\Omega_{ccb}$ and $\Omega_{cbb}$ baryons are calculated with the aid of this above extracted relation.

\subsection{Ground state masses of $\Omega_{ccb}$ and $\Omega_{cbb}$ baryons}

\begin{figure*}
	\begin{center}
	\vspace{-3cm}
		\includegraphics[scale=0.35]{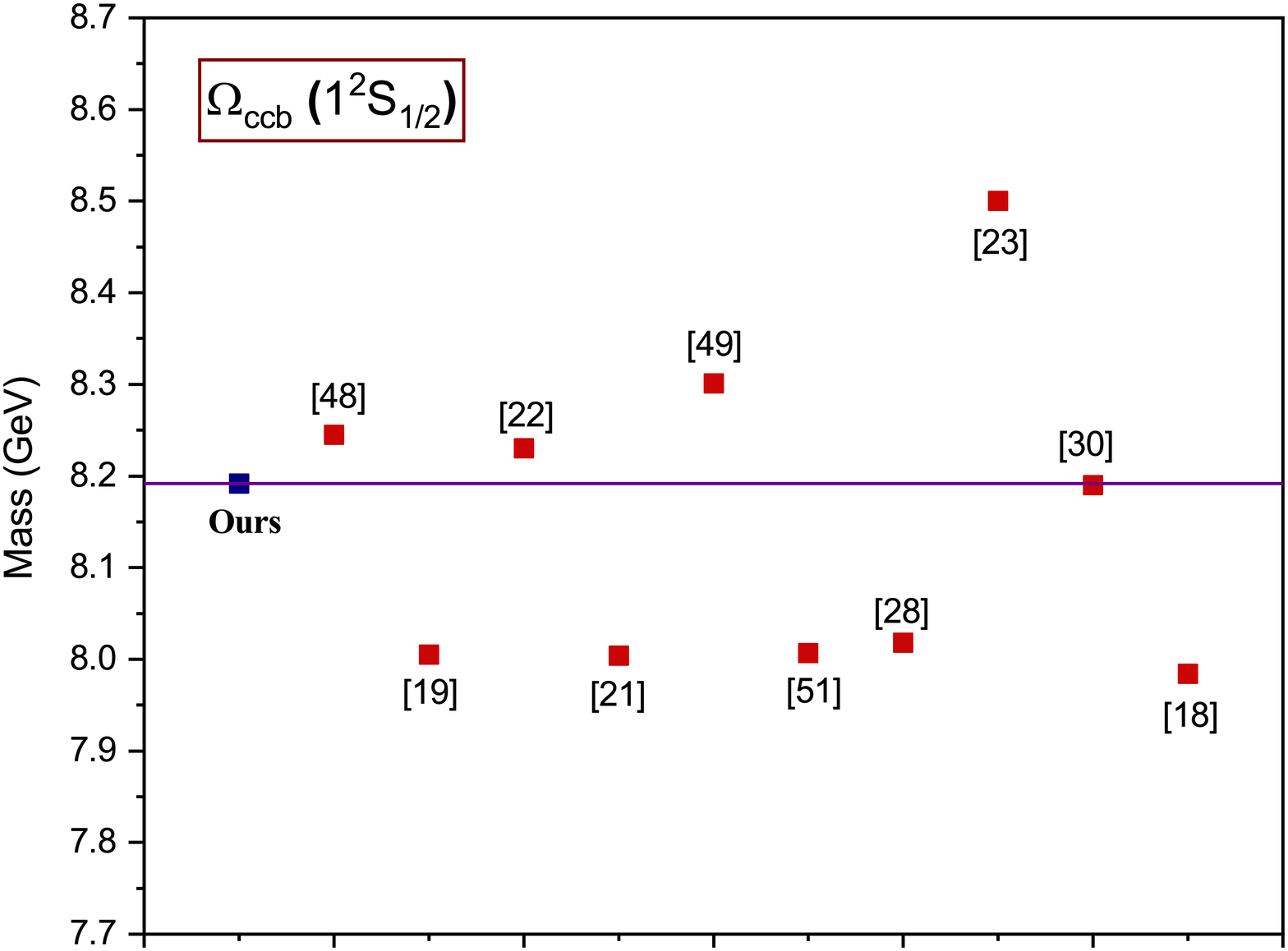}
		\vspace{-1cm} 
		\caption{Comparison between the computed ground state $J^{P} = \frac{1}{2}^{+}$  mass of $\Omega_{ccb}$ baryon with other theoretical approaches.} \label{fig:a}
	\end{center}
\end{figure*}

\begin{figure}
	\centering
	\includegraphics[scale=0.35]{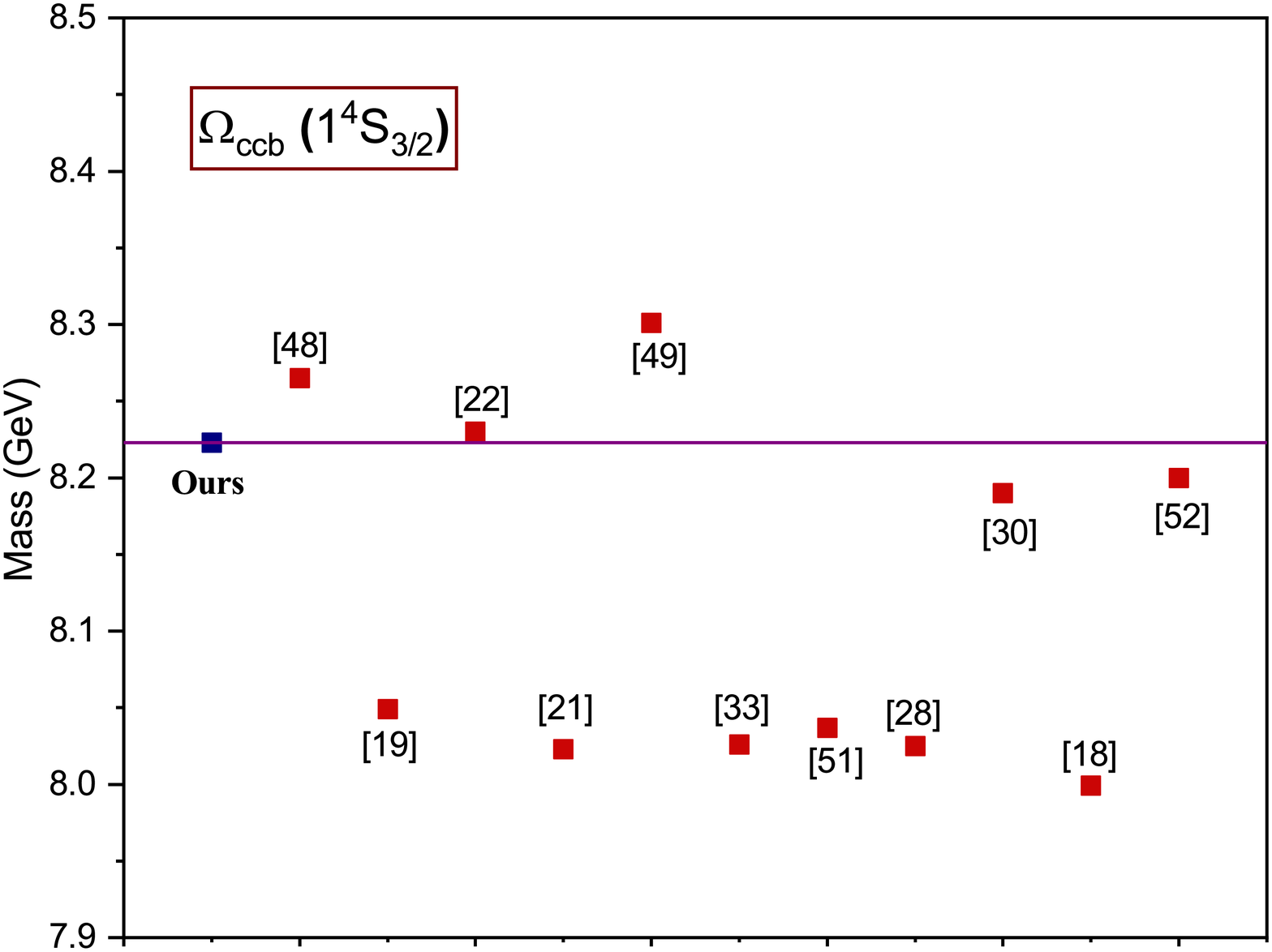}
\vspace{-1cm}
	\caption{\label{fig:b}{Comparison between the computed ground state $J^{P} = \frac{3}{2}^{+}$  mass of $\Omega_{ccb}$ baryon with other theoretical approaches.}}
\end{figure}

\begin{figure}
	\centering
	\includegraphics[scale=0.35]{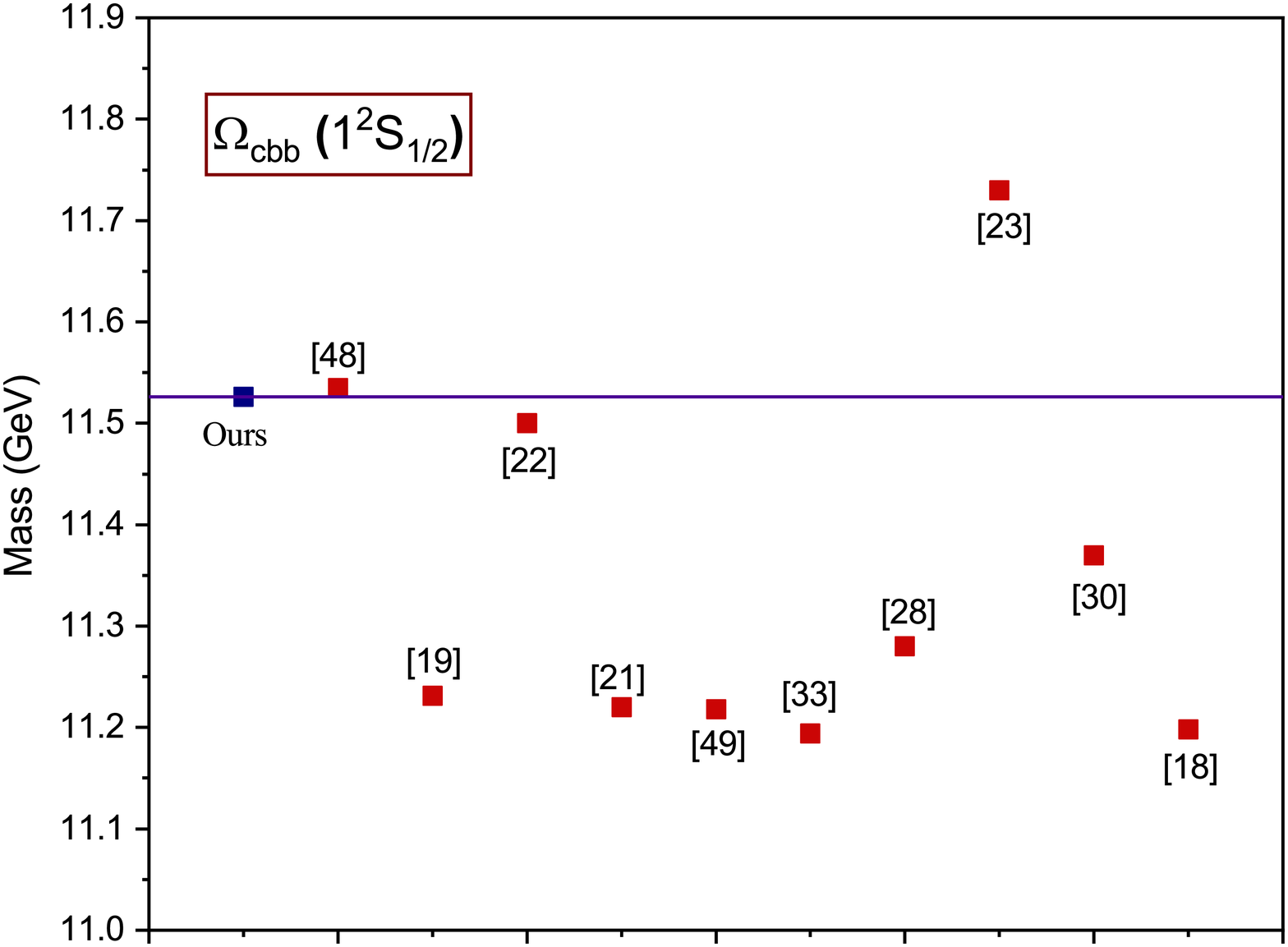}
	\vspace{-2.3cm}
	\caption{\label{fig:c}{Comparison between the computed ground state $J^{P} = \frac{1}{2}^{+}$ mass of $\Omega_{cbb}$ baryon with other theoretical approaches.}}
\end{figure}

\begin{figure}
	\centering
	\includegraphics[scale=0.35]{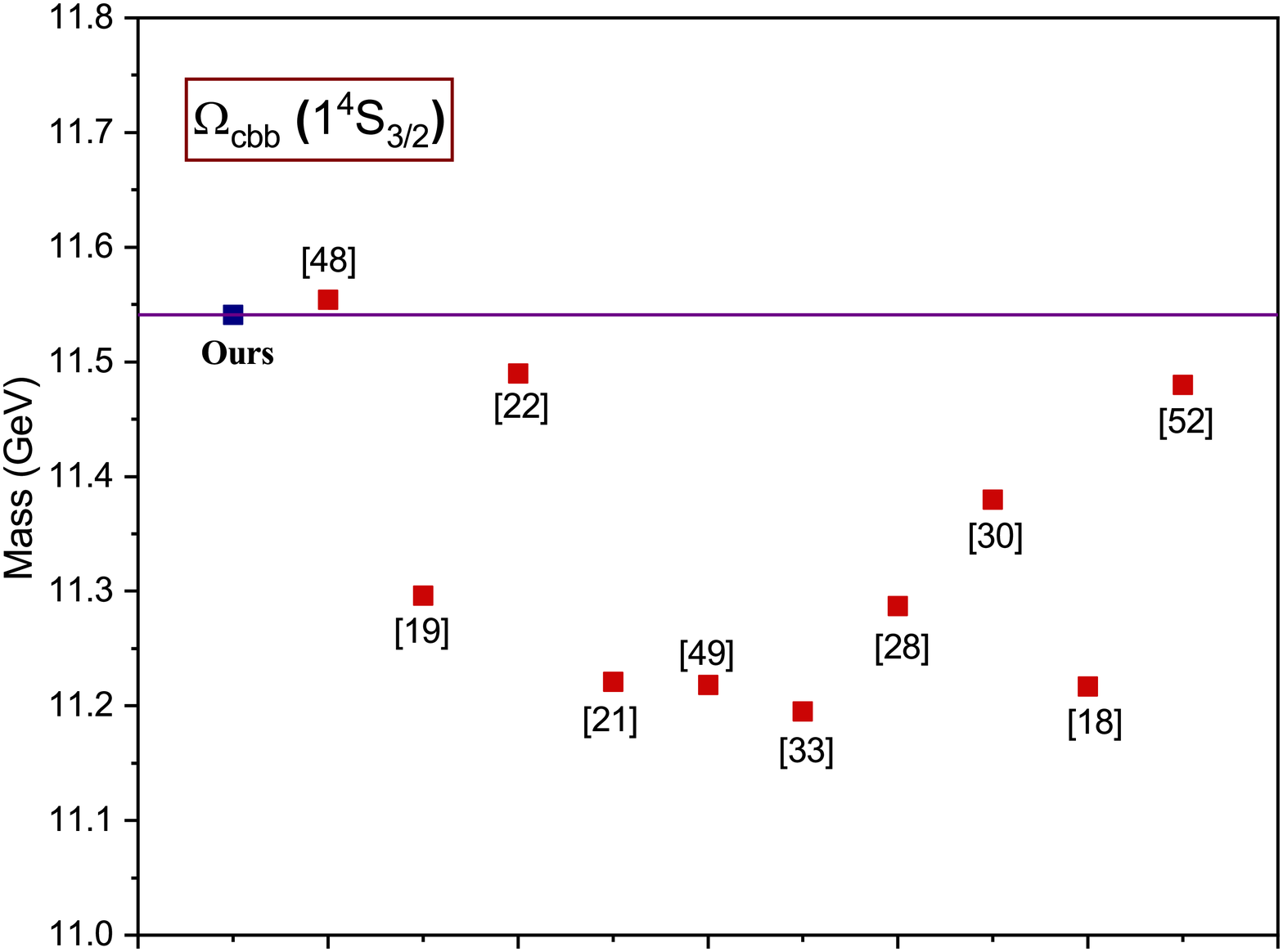}
	\vspace{-2cm}
	\caption{\label{fig:d}{Comparison between the computed ground state $J^{P} = \frac{3}{2}^{+}$  mass of $\Omega_{cbb}$ baryon with other theoretical approaches.}}
\end{figure}

In this section, we evaluate the ground state $\frac{1}{2}^{+}$ and $\frac{3}{2}^{+}$ masses of unseen triply heavy $\Omega_{ccb}$ and $\Omega_{cbb}$ baryons using Eq. (\ref{eq:7}). With the quark composition of the $\Omega_{ccb}$ baryon which \textbf{consists of} two charm quarks ($c$) and one bottom quark ($b$), we put $p = n(u$ or $d)$, $q=c$, $r=b$, and $k=n(u$ or $d)$ in Eq. (\ref{eq:7}), and after simplifying we obtain a quadratic relation in terms of different flavors of baryon masses which is expressed as, 


\begin{eqnarray}
		\label{eq:8} \nonumber
		\frac{4M^{2}_{\Sigma_{b}}M^{2}_{\Omega_{ccb}}}{(4M^{2}_{\Xi_{bc}}-M^{2}_{\Sigma_{b}}-M^{2}_{\Omega_{ccb}})} - (4M^{2}_{\Xi_{bc}}-M^{2}_{\Sigma_{b}}-M^{2}_{\Omega_{ccb}}) \\  \hspace{-1.5cm}
	= \sqrt{M^{2}_{\Sigma_{b}}(M^{2}_{\Sigma_{b}}-2M^{2}_{\Omega_{ccb}}-8M^{2}_{\Xi_{bc}}) + M^{2}_{\Omega_{ccb}}(M^{2}_{\Omega_{ccb}}-M^{2}_{\Xi_{bc}}) + 16 M^{2}_{\Xi_{bc}}}
\end{eqnarray}
similarly, for $\Omega_{cbb}$ baryon which is made up of two bottom quarks and one charm quark, we put $p = n(u$ or $d)$, $q=b$, $r=c$, and $k=n(u$ or $d)$ in Eq. (\ref{eq:7}), we have

\begin{eqnarray}
	\label{eq:9} \nonumber
	\frac{4M^{2}_{\Sigma_{c}}M^{2}_{\Omega_{cbb}}}{(4M^{2}_{\Xi_{bc}}-M^{2}_{\Sigma_{c}}-M^{2}_{\Omega_{cbb}})} - (4M^{2}_{\Xi_{bc}}-M^{2}_{\Sigma_{c}}-M^{2}_{\Omega_{cbb}}) \\  \hspace{-1.5cm}
	= \sqrt{M^{2}_{\Sigma_{c}}(M^{2}_{\Sigma_{c}}-2M^{2}_{\Omega_{cbb}}-8M^{2}_{\Xi_{bc}}) + M^{2}_{\Omega_{cbb}}(M^{2}_{\Omega_{cbb}}-M^{2}_{\Xi_{bc}}) + 16 M^{2}_{\Xi_{bc}}}
\end{eqnarray} 
\\
after inserting the masses of $\Sigma_{b}$ and $\Sigma_{c}$ from PDG \cite{PDG}, and $\Xi_{bc}$ calculated in the previous work \cite{physica} for $J^{P}=\frac{3}{2}^{+}$, into Eqs. (\ref{eq:8}) and (\ref{eq:9}), we \textbf{get the} ground state masses of $\Omega_{ccb}$ and $\Omega_{cbb}$ as 8.223 GeV and 11.541 GeV respectively for $J^{P}=\frac{3}{2}^{+}$. Similarly we can obtain the masses for $J^{P}=\frac{1}{2}^{+}$ as well. The calculated results for ground state masses of $\Omega_{ccb}$ and $\Omega_{bbc}$ baryons are shown in table \ref{tab:1} along with the predicted outcomes of various theoretical approaches. Also, we have shown graphically the comparison of our obtained ground state masses with the results of other theories in Figs \ref{fig:a}-\ref{fig:d}.


\begingroup
\setlength{\tabcolsep}{14pt}
\begin{table*}
	\centering
\caption{\label{tab:1} Ground state $J^{P} = \frac{1}{2}^{+}$ and $\frac{3}{2}^{+}$ masses of $\Omega_{ccb}$ and $\Omega_{cbb}$ baryons (in GeV).}
\hspace{1.5cm}

\begin{tabular}{lllllllllllllc} 
	\hline
	&\multicolumn{2}{c}{$\Omega_{ccb}$} &\multicolumn{2}{c}{$\Omega_{cbb}$} \\
\hline		  
$J^{P}$	&\multicolumn{1}{c}{$\frac{1}{2}^{+}$} &\multicolumn{1}{c}{$\frac{3}{2}^{+}$}&\multicolumn{1}{c}{$\frac{1}{2}^{+}$} 	&\multicolumn{1}{c}{$\frac{3}{2}^{+}$} \\
\hline
Our & 8.192 &8.223 &11.526 &11.541\\
\cite{Roberts2008} &8.245 &8.265 &11.535 &11.554 \\
\cite{Shah2018}& 8.005 &8.049 &11.231 &11.296 \\
\cite{Z.G.Wang2012}	& 8.230 &8.230 &11.500 & 11.490 \\
\cite{G.Yang2020} & 8.004 &8.023 &11.220 &11.221 \\
\cite{Serafin2018} & 8.301 &8.301 &11.218 &11.218 \\
\cite{Padmanath2018} & 8.005 &8.026 &11.194 &11.195 \\
\cite{Z.S.Brown2014} &8.007 &8.037 &11.218 &11.229 \\
\cite{A. P. Martynenko2008} &8.018 &8.025 &11.280  &11.287 \\
\cite{T. M. Aliev2013} &8.500 & &11.730 \\
\cite{P.-L. Yin2019} &8.190 &8.190 &11.370 &11.380\\
\cite{Faustov2022} &7.984 &7.999 &11.198 & 11.217 \\
\cite{J.D.Bjorken} & &8.200 & &11.480 \\
\hline
\end{tabular}
\end{table*}
	\normalsize
\endgroup

\vspace{-0.3cm}

\subsection{Masses in the $(J,M^{2})$ plane}

In this section, we determine the orbitally excited state masses of $\Omega_{ccb}$ and $\Omega_{cbb}$ baryons after calculating their ground state masses. The Regge slopes ($\alpha^{'}$) for  $\frac{1}{2}^{+}$ and $\frac{3}{2}^{+}$ trajectories are evaluated using the relation  (\ref{eq:4}). For instance, to extract the value of Regge slope for $\Omega_{ccb}$ ($\alpha^{'*}_{\Omega_{ccb}}$) for $\frac{3}{2}^{+}$ trajectory, again we insert $p=u$, $q=c$, and $r=b$ into Eq. (\ref{eq:4}) and we get,
\begin{eqnarray}
	\label{eq:10} \nonumber
	\frac{\alpha^{'*}_{\Omega_{ccb}}}{\alpha^{'*}_{\Sigma_{b}}}&=&\frac{1}{2M^{2}_{\Omega_{ccb}}}\times[(4M^{2}_{\Xi_{bc}}-M^{2}_{\Sigma_{b}}-M^{2}_{\Omega_{ccb}}) \\
	&+&\sqrt{{{(4M^{2}_{\Xi_{bc}}-M^{2}_{\Sigma_{b}}-M^{2}_{\Omega_{ccb}}})^2}-4M^{2}_{\Sigma_{b}}M^{2}_{\Omega_{ccb}}}],
\end{eqnarray}
\\
by inserting the masses of $\Sigma_{b}$ from PDG \cite{PDG}, $\Xi_{bc}$ from \cite{physica}, and $\Omega_{ccb}$ (calculated above) into Eq. (\ref{eq:11}), we obtain the value of $\alpha^{'*}_{\Omega_{ccb}}$ as 0.2134 GeV$^{-2}$ for $\frac{3}{2}^{+}$ trajectory. Here we have taken $\alpha^{'*}_{\Sigma_{b}}$ = 0.2906 GeV$^{-2}$ from our previous work \cite{Juhibottom}. In the same manner we can obtain the value of  $\alpha^{'}_{\Omega_{ccb}}$ = 0.1978 GeV$^{-2}$ for $\frac{1}{2}^{+}$ trajectory. 

Similarly, we can get the expression to calculate the Regge slope for $\Omega_{ccb}$ baryon as well by inserting the values of $p, q,$ and $r$ in the relation (\ref{eq:4}) \textbf{according to} its quark composition.  The derived relation is expressed as, 
\begin{eqnarray}
	\label{eq:s} \nonumber
	\frac{\alpha^{'*}_{\Omega_{cbb}}}{\alpha^{'*}_{\Sigma_{c}}}&=&\frac{1}{2M^{2}_{\Omega_{cbb}}}\times[(4M^{2}_{\Xi_{bc}}-M^{2}_{\Sigma_{c}}-M^{2}_{\Omega_{cbb}}) \\
	&+&\sqrt{{{(4M^{2}_{\Xi_{bc}}-M^{2}_{\Sigma_{c}}-M^{2}_{\Omega_{cbb}}})^2}-4M^{2}_{\Sigma_{c}}M^{2}_{\Omega_{cbb}}}],
\end{eqnarray}
again putting the masses of $\Sigma_{b}$ from PDG \cite{PDG}, $\Xi_{bc}$ from \cite{physica}, and $\Omega_{cbb}$ (calculated above) in the above relation, and after simplifying we can have the value of  $\alpha^{'*}_{\Omega_{cbb}}$ as 0.1279 GeV$^{-2}$ for $\frac{3}{2}^{+}$ trajectory. Here also, the $\alpha^{'*}_{\Sigma_{c}}$ = 0.5688 GeV$^{-2}$ is calculated in the same manner as $\alpha^{'*}_{\Sigma_{b}}$.  Also, for $\frac{1}{2}^{+}$ trajectory we have $\alpha^{'}_{\Omega_{cbb}}$ = 0.1188 GeV$^{-2}$.

Now, another expression \textbf{can be} obtained using the Eq. (\ref{eq:1}) in terms of Regge slope and masses, which is expressed as,

\begin{equation}
	\label{eq:11}
	M_{J+1} = \sqrt{M_{J}^{2}+\frac{1}{\alpha^{'}}} .
\end{equation}

With the aid of the above relation and the values of Regge slopes, $\alpha^{'}_{\Omega_{ccb}}$ and $\alpha^{'}_{\Omega_{bbc}}$ extracted above, we can calculate the orbitally excited \textbf{state} masses for triply heavy $\Omega$ baryons for natural parity states ( $J^{P}$ = $1/2^{+}$, $3/2^{-}$, $5/2^{+}$, ....) and unnatural parity states ( $J^{P}$ = $3/2^{+}$, $5/2^{-}$, $7/2^{+}$, ....) lying on that Regge trajectories.
The calculated results are shown in tables \ref{tab:2} and \ref{tab:3} along with the predictions of other theoretical models.

\subsection{Masses in the $(n,M^{2})$ plane}

\begingroup
\setlength{\tabcolsep}{10pt}
\begin{table*}
	\centering
	\caption{\label{tab:2}Excited state masses of $\Omega_{ccb}$ baryon in the ($J,M^{2}$) plane (in GeV).}
	\hspace{1.5cm}
	
	\begin{tabular}{lllllllllllllc} 
	\hline
	States & Our & \cite{Roberts2008} & \cite{G.Yang2020} & \cite{Shah2018} &\cite{Serafin2018} & \cite{Faustov2022}\\
	\textit{$N^{2S+1}L_{J}$} \\
	\hline
		$1^{2}S_{\frac{1}{2}}$ & 8.192 & 8.245 &8.004 &8.005 &8.301 &7.984\\
	$1^{2}P_{\frac{3}{2}}$ &8.495 & 8.420 & & & &8.262 \\
	$1^{2}D_{\frac{5}{2}}$ &8.787 &8.568 & & & &8.473 \\
	$1^{2}F_{\frac{7}{2}}$ &9.070 \\
	$1^{2}G_{\frac{9}{2}}$ &9.344 \\
	$1^{2}H_{\frac{11}{2}}$&9.611 \\
	\noalign{\smallskip}

	$1^{4}S_{\frac{3}{2}}$ & 8.223 & 8.265 &8.023 &8.049 &8.301 &7.999\\
	$1^{4}P_{\frac{5}{2}}$ & 8.503 & 8.432 &8.311 &8.365 &8.491 &8.267\\
	$1^{4}D_{\frac{7}{2}}$ & 8.774 & 8.568 &8.538 &8.780 &8.647 &8.473\\
	$1^{4}F_{\frac{9}{2}}$ & 9.037 & & &9.184\\
	$1^{4}G_{\frac{11}{2}}$& 9.293 & \\
	$1^{4}H_{\frac{13}{2}}$& 9.542 & \\
	\hline
	\end{tabular}
\end{table*}
\normalsize
\endgroup

\begingroup
\setlength{\tabcolsep}{9pt}
\begin{table*}
	\centering
	\caption{\label{tab:3}Excited state masses of $\Omega_{cbb}$ baryon in the ($J,M^{2}$) plane (in GeV).}
	\hspace{1.5cm}
	
	\begin{tabular}{llllllllllllll} 
		\hline
		
	States & Our & \cite{Roberts2008} & \cite{G.Yang2020} & \cite{Zalak2018} & \cite{Faustov2022}& \cite{Z.G.Wang2012} \\
	\textit{$N^{2S+1}L_{J}$} \\
	\hline
	$1^{2}S_{\frac{1}{2}}$ & 11.526 & 11.535 &11.200 &11.231 &11.198 &11.500\\
	$1^{2}P_{\frac{3}{2}}$ & 11.885 &11.711 & & &11.424\\
	$1^{2}D_{\frac{5}{2}}$ & 12.234 &11.823 & & &11.632\\
	$1^{2}F_{\frac{7}{2}}$ & 12.573\\
	$1^{2}G_{\frac{9}{2}}$ & 12.903\\
	$1^{2}H_{\frac{11}{2}}$& 13.225\\
	\noalign{\smallskip}
	$1^{4}S_{\frac{3}{2}}$ & 11.541 & 11.554 &11.221 &11.296 &11.217 &11.490\\
	$1^{4}P_{\frac{5}{2}}$ & 11.886 & 11.762 &11.569 &11.558 &11.543 &\\
	$1^{4}D_{\frac{7}{2}}$ & 12.220 & 11.810 &11.688 &11.909 &11.635 &\\
	$1^{4}F_{\frac{9}{2}}$ & 12.546 & & & 12.249\\
	$1^{4}G_{\frac{11}{2}}$& 12.863 & \\
	$1^{4}H_{\frac{13}{2}}$& 13.173 & \\
		\hline
	\end{tabular}
\end{table*}
\normalsize
\endgroup

After computing the excited orbital states of $\Omega_{ccb}$ and $\Omega_{cbb}$ baryons using the derived relations in the $(J,M^{2})$ plane, in this section we extend our theory to determine the radial excitations of these baryons in the $(n,M^{2})$ plane. The similar equation for the linear Regge trajectories in the ($n,M^{2}$) plane can be expressed as,
\begin{equation}
	\label{eq:12}
	n = \beta_{0} + \beta M^{2},
\end{equation}
where $n$ = 1, 2, 3.... is the principal quantum number, $\beta_{0}$, and $\beta$ are the intercept and slope of the trajectories. 
Since the baryon multiplets lying on the Regge line have the same Regge parameters \cite{Wei2008}, hence to obtain the radially excited state masses firstly we evaluate the values of $\beta$ and $\beta_{0}$ for $S$, $P$, and $D$-waves.  
For $\Omega_{ccb}$ baryon, with the help of the slope equation, we can have  $\beta_{(S)} = 1/(M^{2}_{\Omega_{ccb}(2S)}-M^{2}_{\Omega_{ccb}(1S)})$, where $M_{\Omega_{ccb}(1S)}$ = 8.192 GeV (calculated above). Now since due to the unavailability of experimental masses of triply heavy baryons, to determine the values of Regge parameters we have taken the theoretical predictions as input masses. Hence for $n=2$ the $M_{\Omega_{ccb}(2S)}$ = 8.621 GeV is taken from \cite{Shah2018}. We get $\beta_{(S)}$ = 0.1386 GeV$^{-2}$ for $S$-states for the $1/2^{+}$ trajectory.
Now from Eq. (\ref{eq:12}) we can write,
\begin{eqnarray}
	\label{eq:13}\nonumber
	1 &=& \beta_{0(S)} + \beta_{(S)} M^{2}_{\Omega_{ccb}(1S)},\\ 
	2 &=& \beta_{0(S)} + \beta_{(S)} M^{2}_{\Omega_{ccb}(2S)},
\end{eqnarray}
Using the above relations we can determine the value for Regge intercept as $\beta_{0(S)}$ = -8.3041 for spin $S$ = 1/2. Similarly, we can express these relations for $P$ and $D$-wave as,
\begin{eqnarray}
	\label{eq:14} \nonumber
	1 &=& \beta_{0(P)} + \beta_{(P)} M^{2}_{\Omega_{ccb}(1P)},\\  \nonumber
	2 &=& \beta_{0(P)} + \beta_{(P)} M^{2}_{\Omega_{ccb}(2P)},\\
	1 &=& \beta_{0(D)} + \beta_{(D)} M^{2}_{\Omega_{ccb}(1D)},\\ \nonumber
	2 &=& \beta_{0(D)} + \beta_{(D)} M^{2}_{\Omega_{ccb}(2D)},
\end{eqnarray}

Now, using the above relations we can extract the values of Regge parameters for $\Omega_{ccb}$ baryon for each Regge line and with the aid of these evaluated Regge slopes and intercepts, the radially excited state masses for $n=3,4,5..$ are calculated. The same procedure is follows to  obtain the mass spectra  for $\Omega_{cbb}$ baryon also. The predicted numerical values are shown in Tables \ref{tab:4} and \ref{tab:5} along with the outcomes of other theoretical models.


\begingroup
\setlength{\tabcolsep}{9pt}
\begin{table*}
	\centering
	\caption{\label{tab:4}Excited state masses of $\Omega_{ccb}$ baryon in the ($n,M^{2}$) plane. The numbers in the boldfaced are the inputs taken from \cite{Shah2018} (in GeV).}
	\hspace{1.5cm}
	
	\begin{tabular}{lllllllllllllc} 
		\hline
		
	States & Our & \cite{Roberts2008} & \cite{G.Yang2020} & \cite{Serafin2018} & \cite{Faustov2022} & \cite{Silvestre1996}\\
	\textit{$N^{2S+1}L_{J}$} \\
	\hline
	$1^{2}S_{1/2}$ &8.192 &8.245 &8.004 &8.301 &7.984 &8.019\\
	$2^{2}S_{1/2}$ &\textbf{8.621} &8.537 &8.455 &8.600 &8.361 &8.450\\
	$3^{2}S_{1/2}$ &9.030\\
	$4^{2}S_{1/2}$ &9.420\\
	$5^{2}S_{1/2}$ &9.796\\
	$6^{2}S_{1/2}$ &10.157\\
	
	\noalign{\smallskip}\noalign{\smallskip} 
	
	$1^{4}S_{3/2}$ &8.223 &8.265 &8.023 &8.301 &7.999 &8.056\\
	$2^{4}S_{3/2}$ &\textbf{8.637} &8.553 &8.468 &8.600 &8.366 &8.465\\
	$3^{4}S_{3/2}$ &9.032\\
	$4^{4}S_{3/2}$ &9.410\\
	$5^{4}S_{3/2}$ &9.774\\
	$6^{4}S_{3/2}$ &10.125\\
	
	\noalign{\smallskip}\noalign{\smallskip} 
	
	$1^{4}P_{5/2}$  &8.503 &8.432 &8.311 &8.491 &8.267 &8.331\\
	$2^{4}P_{5/2}$ &\textbf{8.955} & &8.667 & & &8.589\\
	$3^{4}P_{5/2}$ &9.385\\
	$4^{4}P_{5/2}$ &9.797\\
	$5^{4}P_{5/2}$ &10.191\\
	
	\noalign{\smallskip}\noalign{\smallskip}    
	
	$1^{4}D_{7/2}$ &8.774 &8.568 & 8.538 &8.647 &8.473 &8.528\\
	$2^{4}D_{7/2}$ &\textbf{9.368} & &8.839 & & &8.762\\
	$3^{4}D_{7/2}$ &9.926\\
	$4^{4}D_{7/2}$ &10.455\\
	$5^{4}D_{7/2}$ &10.959\\
	
		\hline
	\end{tabular}
\end{table*}
\normalsize
\endgroup

\begingroup
\setlength{\tabcolsep}{9pt}
\begin{table*}
	\centering
	\caption{\label{tab:5}Masses for $\Omega_{cbb}$ baryon in the ($n,M^{2}$) plane. The numbers in the boldfaced are the inputs taken from \cite{Zalak2018} (in GeV).}
	\hspace{1.5cm}
	
	\begin{tabular}{lllllllllllllc} 
		\hline
		
		States & Our & \cite{Roberts2008} & \cite{G.Yang2020} & \cite{Serafin2018} & \cite{Faustov2022} & \cite{Silvestre1996}  \\
	\textit{$N^{2S+1}L_{J}$} \\
	\hline
	$1^{2}S_{1/2}$ &11.526 & 11.535 &11.200 &11.218 &11.198 &11.217\\
	$2^{2}S_{1/2}$ &\textbf{11.757} &11.787 &11.607 &11.585 &11.507 &11.625\\
	$3^{2}S_{1/2}$ &11.984\\
	$4^{2}S_{1/2}$ &12.206\\
	$5^{2}S_{1/2}$ &12.424\\
	$6^{2}S_{1/2}$ &12.639\\
	
	\noalign{\smallskip}\noalign{\smallskip} 
	
	$1^{4}S_{3/2}$ &11.541 &11.554 &11.221 &11.218 &11.217 &11.251\\
	$2^{4}S_{3/2}$ &\textbf{11.779} &11.798 &11.622 &11.585 &11.515 &11.643\\
	$3^{4}S_{3/2}$ &12.012\\
	$4^{4}S_{3/2}$ &12.241\\
	$5^{4}S_{3/2}$ &12.466\\
	$6^{4}S_{3/2}$ &12.686\\
	
	\noalign{\smallskip}\noalign{\smallskip} 
	
	$1^{4}P_{5/2}$ &11.886 &11.762 &11.569 &11.601 &11.543 &11.598\\
	$2^{4}P_{5/2}$ &\textbf{12.049} & &11.888 & & &11.899\\
	$3^{4}P_{5/2}$ &12.210\\
	$4^{4}P_{5/2}$ &12.369\\
	$5^{4}P_{5/2}$ &12.525\\
	
	\noalign{\smallskip}\noalign{\smallskip}    
	
	$1^{4}D_{7/2}$ &12.220 &11.810 &11.688 &11.626 &11.635 &11.718\\
	$2^{4}D_{7/2}$ &\textbf{12.396} & &11.963 & & &11.986\\
	$3^{4}D_{7/2}$ &12.569\\
	$4^{4}D_{7/2}$ &12.741\\
	$5^{4}D_{7/2}$ &12.910\\
	
		\hline
	\end{tabular}
\end{table*}
\normalsize
\endgroup

\section{Magnetic moment}
The study of electromagnetic properties of baryons is a significant topic for theoretical as well as experimental work. The Magnetic moment is an intrinsic property that aids in understand the shape and other dynamics of transitions in decay modes. The magnetic moment expression can be obtained by operating the spin-flavour wave function to the $z$-component of the magnetic moment operator.
The generalised form of magnetic moment is given as \cite{,B. Patel20009,RohitDhir2021}.

\begin{equation}
	\mu_{B} = \sum_{i} \langle \phi_{sf}|\mu_{iz}| \phi_{sf} \rangle,
\end{equation}
where, $i$ = $c$ or $b$. $\phi_{sf}$ represents the spin-flavour wave function of the baryons. 
The magnetic moment of the individual quark appears as,
\begin{equation}
    \mu_{iz} = \dfrac{e_{i}}{2m_{i}^{eff}}\sigma_{iz},
\end{equation}
here $e_{i}$ is the quark charge, $\sigma_{iz}$ is the spin of the respective constituent quark    
corresponding to the spin flavour wave-function of the baryonic state and $m_{i}^{eff}$ is the effective mass of the constituent quarks \cite{AmeeIJMPA2022,RohitDhir2021}. Since we have not used the constituent quark masses in our formalism to predict the mass spectra. Here we have taken the values of these quark masses, $m_{c}$ = 1.275 GeV and $m_{b}$ = 4.250 GeV, constituent masses of the charm and bottom quark respectively,  from Particle data group (PDG) \cite{PDG}.  
Using the above equations we have calculated the  ground state magnetic moments of $\Omega_{ccb}$ and $\Omega_{cbb}$ baryons. The final spin-flavor wave function and the evaluated results of magnetic moment along with the predictions of other theoretical approaches are shown in table \ref{tab:MM}.

\begin{table*}
	\centering
	\caption{\label{tab:MM}Magnetic moments (in $\mu_{N}$) of $J^{P}$ = $\frac{1}{2}^{+}$ and $\frac{3}{2}^{+}$ of $\Omega_{ccb}$ and $\Omega_{cbb}$ baryons.}
	\hspace{1.5cm}
	
	\begin{tabular}{ccccccccccccccccccc} 
		\hline
		
		Spin & Baryon & $\sigma_{iz}$ & $\mu$ (Ours) & \cite{RohitDhir2021} &\cite{Bernotas} & \cite{B. Patel20009} \\
		\hline
		$\frac{1}{2}^{+}$ & $\Omega_{ccb}$ & $\frac{4}{3}$ $\mu_{c}-\frac{1}{3} \mu_{b}$ & 0.5632 & 0.5261 & 0.540 & 0.502 \\
			\noalign{\smallskip}
		$\frac{3}{2}^{+}$ & $\Omega_{ccb}^{*}$ &  $\mu_{b}+2\mu_{c}$ & 0.7503 & 0.6952 &  0.720 &0.651\\
			\noalign{\smallskip}
		$\frac{1}{2}^{+}$ & $\Omega_{cbb}$ & $\frac{4}{3} \mu_{b}+\frac{1}{3} \mu_{c}$ & -0.2218 & -0.2096 &-0.210 & -0.203 \\
			\noalign{\smallskip}
		$\frac{3}{2}^{+}$ & $\Omega_{cbb}^{*}$ &  $2\mu_{b}+\mu_{c}$ & 0.2908 &0.2558 & 0.270 & 0.216 \\
			\noalign{\smallskip}
			\hline
	\end{tabular}
\end{table*}

\section{Radiative decay width}

Radiative decay is the process of $\gamma$ emission  occurring as the spin state changes from $\frac{3}{2}$ to $\frac{1}{2}$. For the better understanding of the intrinsic properties of heavy baryons, the experimental determination of the radiative decay width is important. The expression for radiative decay width is given by \cite{RohitDhir2021}, 


\begin{equation}
	\Gamma_{R} = \dfrac{k^{3}}{m_{p}^{2}}\dfrac{2}{2J+1}\dfrac{e^{2}}{4\pi}|\mu_{B\rightarrow B^{'}}|^{2},
\end{equation}
where,
\begin{equation}
	k =\dfrac{M^{2}_{B} - M^{2}_{B^{'}}}{2M_{B}}
\end{equation}
where $k$ is the photon energy, $M_{B}$ and $M_{B^{'}}$ are the masses of initial and final baryon state, respectively. $m_{p}$ is the photon mass, and J is the initial angular momentum. Also, $\mu_{B\rightarrow B^{'}}$ is the transition magnetic moment which is expressed as, 
\begin{equation}
	\mu_{B\rightarrow B^{'}} = \langle \phi_{B} |\mu_{B\rightarrow B^{'}}| \phi_{B^{'}} \rangle
\end{equation}
Here, $\phi_{B}$ and $\phi_{B^{'}}$ are the wave functions of initial and final baryonic states respectively. Table \ref{tab:DR} shows the calculated transition magnetic moment and radiative decay widths of $\Omega_{ccb}$ and $\Omega_{cbb}$ baryons. 

\begin{table*}[h!]
	\centering
	\caption{\label{tab:DR} The Radiative decay widths (in keV) of $\Omega_{ccb}$ and $\Omega_{cbb}$ baryons.}
	\hspace{1.5cm}
	
	\begin{tabular}{ccccccccccccccccccc} 
		\hline
		
		Decay & Wave-function & Transition magnetic & $\Gamma_{R}$ (Ours) & $\Gamma_{R}$ \cite{RohitDhir2021} & $\Gamma_{R}$ \cite{Simonis2013}  \\
		& & moment ($\mu_{N}$)  & (keV) \\
		\hline
		$\Omega_{ccb}^{*+} \rightarrow \Omega_{ccb}^{+} \gamma $ & $\frac{2\sqrt{2}}{3} (\mu_{c}-\mu_{b})$ & 0.441 & 0.0238 & 0.0192 &0.004\\
		
		\noalign{\smallskip}
	
	$\Omega_{cbb}^{*0} \rightarrow \Omega_{cbb}^{+} \gamma $ & $\frac{2\sqrt{2}}{3} (\mu_{b}-\mu_{c})$ & -0.451 & 0.0028 & 0.0282 & 0.005 \\
		\hline
	\end{tabular}
\end{table*}

\section{Results and Discussion}

The spectroscopic study of $\Omega_{ccb}$ and $\Omega_{cbb}$ baryon sectors is successfully obtained in the present work using the Regge phenomenology. We have evaluated the ground and the excited state masses of these triply heavy baryons. With no experimental information available on these systems, we have compared our estimated results with recent theoretical predictions. Table \ref{tab:1}. shows the calculated ground state masses of $\Omega_{ccb}$ and $\Omega_{cbb}$ baryons along with the comparison of the results obtained from different theoretical approaches. Tables \ref{tab:2} and \ref{eq:3} represents the orbitally excited state masses evaluated in the $(J,M^{2})$ plane for these triply heavy baryons. Further, our calculated results for the radially excited states in the $(n,M^{2})$ plane are \textbf{summarised} in tables \ref{tab:4} and \ref{tab:5}. \textbf{The properties like magnetic moment and radiative decay for $\Omega_{cbb}$ baryon have also been calculated. The obtained results along with the comparison of the predictions from other appproaches are shown in table \ref{tab:MM} and \ref{tab:DR}.} 

\subsection{The $\Omega_{ccb}$ baryon sector}
The ground state masses obtained in the present work for $\Omega_{ccb}$ baryon are 8.192 GeV and 8.223 GeV for  $\frac{1}{2}^{+}$ and $\frac{3}{2}^{+}$ states respectively. The ground state masses predicted in the recent studies by the different theoretical approaches vary in the range of 8.0 - 8.5 GeV and our calculated masses fall within this range (shown in table \ref{tab:1}). The estimated masses for $\frac{1}{2}^{+}$ and $\frac{3}{2}^{+}$ states are in well agreement with the predicted masses obtained in the non-relativistic quark model \cite{Roberts2008} and seem to be very close to the results obtained using QCD sum rules \cite{Z.G.Wang2012} with a mass difference of few MeV. In contrast, some studies \cite{Shah2018,G.Yang2020,Z.S.Brown2014,A. P. Martynenko2008} show a quite large mass difference of around 170-200 MeV with our predicted masses. This can also be seen in graphical representation where our calculated masses are presented by "Ours" which lies in between the spreaded results of other theoretical predictions (shown in Figs. \ref{fig:a} and \ref{fig:b}). Further, the calculated masses for orbitally excited states of $\Omega_{ccb}$ baryons are shown in table \ref{tab:2}. Again we compared our results with other theoretical predictions and a general agreement can be seen with the results of some references. The predicted mass for $1^{2}P_{\frac{3}{2}}$ state is close the results of \cite{Roberts2008} and shows a large mass difference with the prediction of \cite{Faustov2022}. For the $1^{4}P_{\frac{5}{2}}$ state, the calculated mass is very close to the result obtained in \cite{Serafin2018} with a mass difference of 12 MeV and also in agreement with the prediction of \cite{Roberts2008}. The results presented in Refs. \cite{G.Yang2020,Shah2018,Faustov2022} are found to be somewhat lower than our obtained mass. Similarly, the masses estimated for $1^{4}D_{\frac{7}{2}}$ and $1^{4}F_{\frac{9}{2}}$ states are close to the results of \cite{Shah2018} and comparatively higher  than the results obtained in Refs. \cite{Roberts2008,G.Yang2020,Faustov2022}. Table \ref{tab:4} shows radially excited state masses obtained in the $(n,M^{2})$ plane. Very few results are available from previous theoretical predictions. The masses reported in the Refs. \cite{Roberts2008,Serafin2018} shows a general agreement with our predicted masses with a mass difference of few MeV. 

\textbf{Further using the obtained results we calculate other properties like the magnetic moment of ground states and the radiative decay width. We compared our results with the predictions of various other theoretical approaches \cite{RohitDhir2021,Bernotas,B. Patel20009} and they are in consistent with our results. 
}

\subsection{The $\Omega_{cbb}$ baryon sector}

Similar behaviour can be seen in the case of $\Omega_{cbb}$ baryon. The ground state masses predicted in  other references vary in the range of  11.2-11.7 GeV and our estimated masses of 11.526 GeV and 11.541 GeV for $\frac{1}{2}^{+}$ and $\frac{3}{2}^{+}$ states respectively lie in this range.  The predicted ground state masses in Refs. \cite{Roberts2008,Z.G.Wang2012,J.D.Bjorken} are very close to our obtained masses with a mass difference of few MeV but the results obtained in other refs. \cite{Zalak2018,G.Yang2020,Serafin2018,A. P. Martynenko2008} shows a bit large mass difference of 0.2-0.3 GeV with our predicted masses as shown in table \ref{tab:1}. Figs \ref{fig:c} and \ref{fig:d} which shows the comparison of our obtained ground state masses with other theoretical results shows a clear picture that our calculated masses fall within the range of the masses obtained from various theories. Tables \ref{tab:3} and \ref{tab:5} represent the orbital and radial excited state masses calculated in the $(J,M^{2})$ and $(n,M^{2})$ planes respectively. We compared our computed results with various theoretical predictions. For low-lying states, the predicted states reported in \cite{Roberts2008} agree with our calculated masses. As we go for higher excited states, quite different results with large uncertainties can be seen in the masses predicted by different theoretical models. There is no clear consensus between the different approaches for the excited state masses of $\Omega_{cbb}$ baryon. The masses reported in the Ref. \cite{Roberts2008,Silvestre1996} agrees well with our predicted masses and the predicted states reported in Refs. \cite{Zalak2018,Faustov2022,G.Yang2020} are lower than ours. 

\textbf{The other properties such as magnetic moment and radiative decay width of are evaluated. The summerized results along with the comparsion from various theoretical predictions are shown in tables \ref{tab:MM} and \ref{tab:DR}. We compared our calculated magnetic moment values with the results obtained in \cite{RohitDhir2021,Bernotas,Simonis2013} and they are in agreement with them. The obtained rediative decay width shows variation with the predictions of \cite{RohitDhir2021,Simonis2013}, it may be due to the large variation in the ground state masses of $\Omega_{cbb}$ baryon.}

\section{Conclusion}

In this paper, we applied the Regge phenomenology for the calculation of the mass spectra of $\Omega_{ccb}$ and $\Omega_{cbb}$ baryons. Different results with large uncertainties are found in different theoretical approaches. The ground state masses of $\Omega_{ccb}$ and $\Omega_{cbb}$ baryons obtained from various studies \textbf{vary} in the range of 8.0-8.5 GeV and 11.2-11.7 GeV respectively and our obtained masses fall within this range. The excited states of these triply heavy \textbf{baryons have been} less explored theoretically. We compared our obtained results with the masses of various theoretical approaches. Here, we reached the conclusion that the low lying masses are in well agreement with the predictions of some theories. However discrepancies are found \textbf{in} excited states.
Since different approaches gives different predictions, hence more comparison of  computations with experimental measurements is required. 
Also, the mass values which are taken as inputs in the calculation are the experimental masses  where available and the theoretical masses which we have calculated in our previous work. \textbf{The model's reliability depend on the availability of experimental data that we have taken as a input mass, hence whatever the mass value predictions we have obtained in the present work follows a general trend with the results of other theoretical models.} \textbf{Also, there are no fitting parameters involved in our formalism, the Regge parameters are calculated from the experimental masses in the present work. We have also seen that the results obtained in some theoretical approaches like hCQM \cite{Shah2018,Zalak2018}, does not follow the general trend. For example, the highest $1P$ state has the lower mass with respect to the lower $1P$ state.}

\textbf{Further, the other properties such as magnetic moment and radiative decay widths have been obtained using the mass value predictions. They also shown a general agreement with other the theoretical predictions, whereas some results shows slight discrepancies.}
Since, there is no experimental data is available related to triply heavy baryons. Hence, our predicted mass values,\textbf{ magnetic moment and decay widths } could be extremely valuable in future experimental searches.

\end{document}